\documentclass[3p,times,twocolumn]{elsarticle}
\usepackage{epsfig}
\usepackage{amssymb}
\usepackage[figuresright]{rotating}
\usepackage{color}

\newlength{\feynwidth} 
\newcommand{\D}{\cdot\cdot\cdot}
\newcommand{\La}{{\Lambda}}
\newcommand{\Si}{{\Sigma}}

\newcommand{\be}{\begin{eqnarray}}
\newcommand{\ee}{\end{eqnarray}}

\begin{document}

\begin{frontmatter}

\title{Implications of an increased $\Lambda$-separation energy of the hypertriton}

 \author[J]{Hoai Le}
 \author[J]{Johann Haidenbauer}
 \author[B,J,C]{Ulf-G. Mei{\ss}ner}
 \author[J]{Andreas Nogga}

 \address[J]{Institute for Advanced Simulation, Institut f{\"u}r Kernphysik and
 J\"ulich Center for Hadron Physics, Forschungszentrum J{\"u}lich, D-52425 J{\"u}lich, Germany}
 \address[B]{Helmholtz Institut f\"ur Strahlen- und Kernphysik and Bethe Center
  for Theoretical Physics, Universit\"at Bonn, D-53115 Bonn, Germany}
 \address[C]{Tbilisi State  University,  0186 Tbilisi, Georgia} 

\begin{abstract}
Stimulated by recent indications that the binding energy of 
the hypertriton could be significantly larger than so far assumed,
requirements of a more strongly bound $^3_\Lambda {\rm H}$ state
for the hyperon-nucleon interaction and consequences for the binding 
energies of $A=4,5$ and $7$ hypernuclei are investigated. 
As basis, a $YN$ potential derived at next-to-leading order in 
chiral effective field theory is employed. Faddeev  and Yakubovsky
equations are solved to obtain the corresponding $3$- and $4$-body 
binding energies, respectively, and the Jacobi no-core shell model is used for 
$^5_\Lambda$He and $^7_\Lambda$Li. 
It is found that the spin-singlet $\Lambda p$ interaction would have to 
be much more attractive which can be, however, accommodated
within the bounds set by the available $\Lambda p$ scattering data. 
The binding energies of the $^4_\Lambda {\rm He}$ hypernucleus 
are predicted to be closer to the empirical values than for 
$YN$ interactions that produce a more weakly bound 
$^3_\Lambda {\rm H}$. The quality of the description of the 
separation energy and excitation spectrum for $^7_\Lambda$Li remains
essentially unchanged. 
\end{abstract}
\begin{keyword}
\PACS 
{13.75.Ev} \sep
{21.80.+a} \sep
{21.30.Fe}
\end{keyword}

\end{frontmatter}

\section{Introduction}

Light hypernuclei play an essential role for testing our 
understanding of the hyperon-nucleon ($YN$) interaction. 
Over the last three decades or so, techniques for treating 
few-body systems have matured to a level that a rigorous 
assessment of sophisticated two-body potentials, including the 
full complexity of $YN$ dynamics like tensor forces or the important
coupling between the $\La N$ and $\Si N$ channels, has become feasible. 
For example, binding energies of $A=3$ and $4$ hypernuclei can be
obtained by solving ``exact'' Faddeev or Yakubovsky equations 
\cite{Miyagawa:1993,Miyagawa:1995,Nogga:2002} based on such $YN$ interactions. 
So-called {\it ab initio} methods like the no-core shell model allow one 
to perform rigorous calculations even for hypernuclei beyond the $s$ shell 
\cite{Bogner:2009bt,Wirth:2014,Wirth:2016,Wirth:2018,Wirth:2019,Liebig:2015kwa,Le:jncsm}
and, so far, studies for hypernuclei up to $^{13}_{\La} {\rm C}$ 
have been reported \cite{Wirth:2019}.  

Of course, for making solid conclusions, it is mandatory that there is likewise 
solid experimental information on the binding energies of hypernuclei. 
Indeed, in recent times, some of the past values have been called into question 
and ``critically revised'' \cite{Botta:2017}. This concerns also the
binding energies of the $A=4$ system where new measurements have been
performed in an attempt to settle the long-standing issue of the large charge 
symmetry breaking (CSB) observed in the binding energies of
the ${^{\, 4}_\La \rm He}$ and ${^{\, 4}_\La \rm H}$ 
hypernuclei \cite{Gal:2016}. The new measurements, performed for the
${^{\, 4}_\La \rm H} \, (0^+)$ state \cite{Esser:2015,Schulz:2016kdc} and the
splitting between the ${^{\, 4}_\La \rm He}$ $0^+$ and $1^+$ levels
\cite{Yamamoto:2015} differ noticeably from the earlier values in the 
literature \cite{Davis:1992}. 
Another binding energy that has been challenged lately 
is that of the hypertriton ${^{\, 3}_\La \rm H}$. Here the value for the separation 
energy, accepted as benchmark for decades, is $E_\Lambda = 0.13 \pm 0.05$~MeV \cite{Juric:1973}, 
while a new measurement by the STAR collaboration suggests a value of $0.41 \pm 0.12$~MeV
\cite{Adam:2019}. This is a quite dramatic increase. 
Actually, there is support for a more tightly bound ${^{\, 3}_\La \rm H}$ by 
recent measurements of the $^3_\Lambda {\rm H}$ lifetime as well. 
Some of the experiments yield values well below that of a free $\Lambda$ 
\cite{Adam:2016,Adamczyk:2018,Puccio:2018,Braun-Munzinger:2019}
which could be a signal for a stronger binding of the hypertriton \cite{Gal:2018}. 

In the present work, we address the consequences of a potentially more strongly bound 
hypertriton. The first question that arises is, of course, which 
modifications of the underlying $YN$ forces are needed in order to achieve a larger binding energy. After all, a correspondingly
modified $YN$ interaction should still be realistic, i.e. it should still
be  in line with existing empirical information on $\La N$ 
and $\Si N$ scattering. 
As a matter of fact, past calculations of the hypertriton within the Faddeev approach 
\cite{Miyagawa:1995,Nogga:2002,Tominaga:2001,Garcilazo:2007,Fujiwara:2008,Haidenbauer:2019} 
have revealed that only some of the $YN$ potentials in the literature lead to a bound hypertriton. For many of the interactions 
considered, it turned out that there was not sufficient 
attraction to support a $\La NN$ bound state
\cite{Miyagawa:1993,Nogga:2002}. 
 
The second interesting question is, what will be the implications
for the $A=4$ system and for heavier hypernuclei. 
Will these be already overbound by a suitably modified $YN$ interaction that supports a larger ${^{\, 3}_\La \rm H}$ binding energy? 
Or does it actually bring the binding 
energy for four-body systems closer to the empirical values? 
Indeed, as reported in Refs.~\cite{Nogga:2002,Haidenbauer:2019},
none of the realistic $YN$ potentials 
\cite{Haidenbauer:2019,Rijken:1999,Haidenbauer:2005,Haidenbauer:2013}  
examined so far in four-body calculations 
yields ${^{\, 4}_\La \rm H}$ (${^{\, 4}_\La \rm He}$) binding energies 
close to the experiment. 
In the exceptional case of the leading order (LO) chiral 
$YN$ interactions \cite{Polinder:2006}, one has to consider that it 
does not really provide a satisfying description of the $\La p$ 
data and that corresponding few-body results are afflicted by a 
sizable cutoff dependence \cite{Gazda:2016}.

\begin{figure}[t]
\begin{center}
\includegraphics[height=100mm]{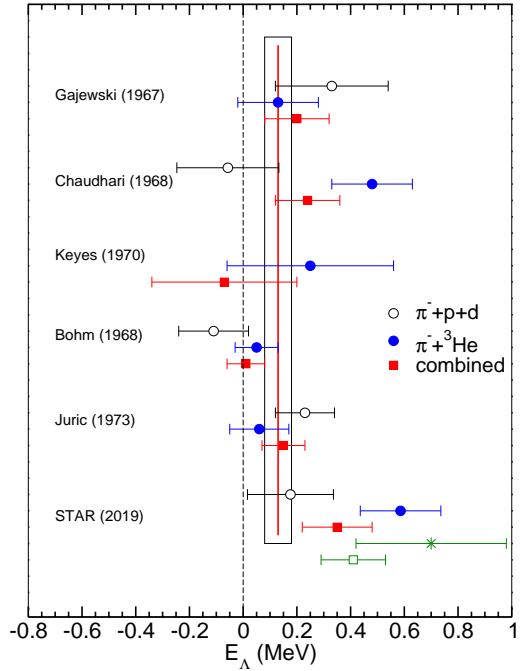}
\caption{
Experimental hypertriton separation energies $E_\La$ from the
literature
\cite{Gajewski:1967,Chaudhari:1968,Keyes:1970,Bohm:1968,Juric:1973,Adam:2019}.
Solid (opaque) circles indicate measurements from the
$\pi^- + {}^3{\rm He}$
($\pi^- + p + d$)
decay channels. Squares indicate combined results. 
The breakdown of the STAR value for $^3_\Lambda {\rm H}$ 
into the two decay channels is based on the preliminary results 
reported in Ref.~\cite{Liu:2019}. Furthermore, the asterisk 
indicates the STAR result for $^3_{\bar\Lambda} {\rm \bar H}$, while
the opaque square represents the combined ($^3_\Lambda {\rm H}$,
$^3_{\bar\Lambda} {\rm \bar H}$) value \cite{Adam:2019}. 
The box indicates the benchmark due to Juri\v c et al.~\cite{Juric:1973},
obtained by combining their own result with the data from 
Bohm et al.~\cite{Bohm:1968}. 
\vspace{-0.7cm}}
\label{fig:F1}
\end{center}
\end{figure}

Before proceeding to the actual calculations, we summarize the
situation concerning the separation energy of the hypertriton. This is done in
Fig.~\ref{fig:F1} where various values from the literature are included
\cite{Gajewski:1967,Chaudhari:1968,Keyes:1970,Bohm:1968,Juric:1973,Adam:2019,Liu:2019}. 
Similar graphical representations have been shown in Refs.~\cite{Achenbach:2018,Adam:2019}. 
One can see that there is quite some variation between the results from different
groups but also between the energies determined from the two decay channels
${^{\, 3}_\La \rm H} \to \pi^- + {}^3{\rm He}$ and
${^{\, 3}_\La \rm H} \to \pi^- + p + d$. Obviously, 
the new STAR measurement is well within the variations of former investigations,
if one leaves the value for the anti-hypertriton separation energy aside.  

\section{Calculation}
\begin{figure}[t]
\begin{center}
\includegraphics[height=115mm]{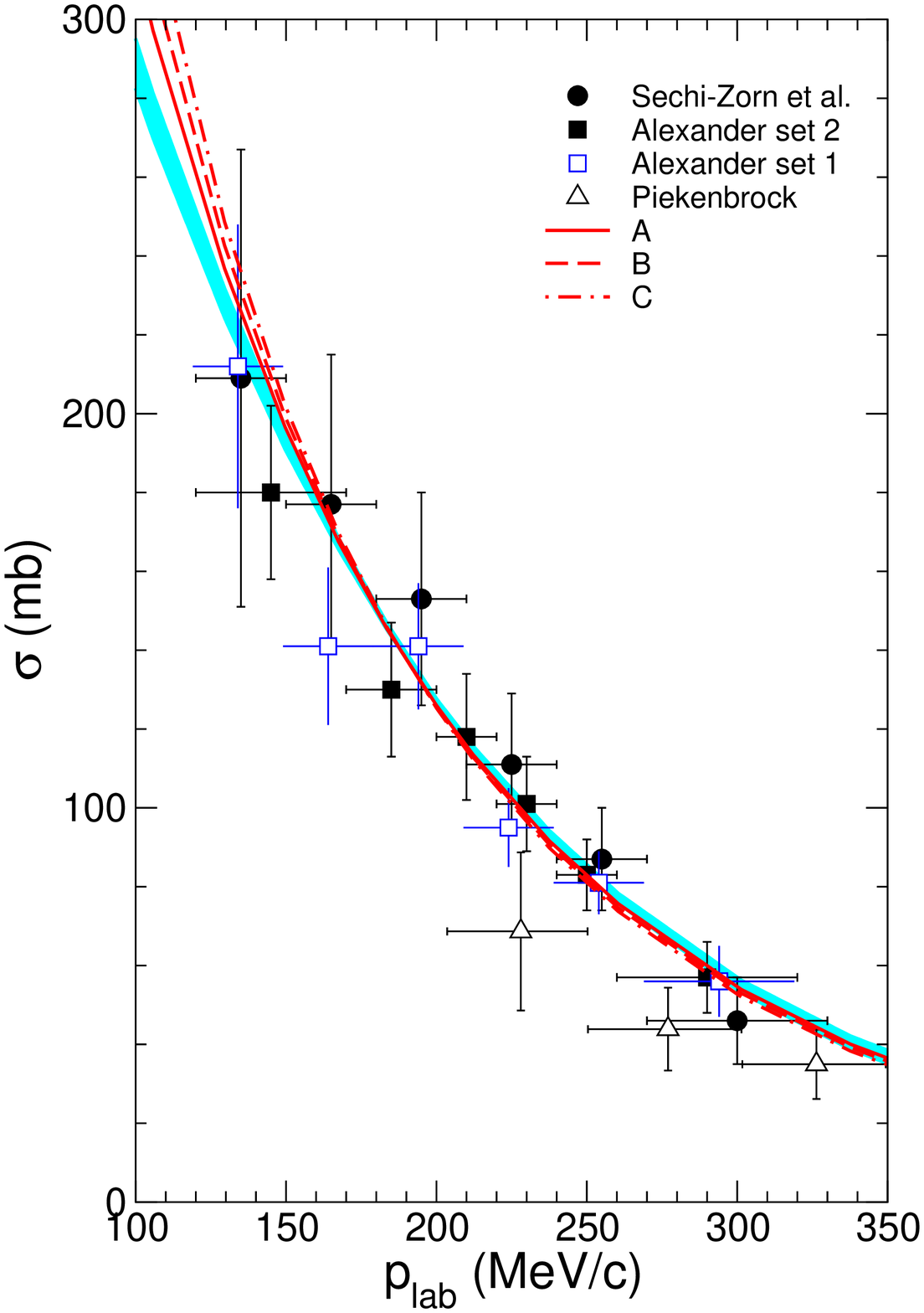}
\vspace{-1.0cm}
\caption{Near-threshold cross section for $\La p$ scattering.
The band represents the result for the $YN$ potential NLO19 
\cite{Haidenbauer:2019} derived within chiral EFT, 
including cutoff variations of $\Lambda = 500-650$~MeV.  
The solid, dashed and dash-dotted
lines corresponding to the fits A, B, and C, see text and Table \ref{tab:T1}. 
The experimental cross sections are taken from Refs.~\cite{Sec68} (filled circles),
\cite{Ale68} (Set 2: filled squares, Set 1: open squares),
\cite{Piekenbrock,Herndon:19672} (open triangles).
\vspace{-0.7cm}}
\label{fig:LN}
\end{center}
\end{figure}

Starting point of the present study is a modern $YN$ interaction derived within 
SU(3) chiral effective field theory (EFT) \cite{Haidenbauer:2013,Haidenbauer:2019}, 
in close analogy to $NN$ forces established in the same framework
\cite{Epelbaum:2005,Epelbaum:2006,Epelbaum:2008}.
In the considered chiral expansion up to next-to-leading order (NLO), 
the $YN$ potential consists of contributions from one- and two-pseudoscalar-meson 
exchange diagrams (involving the Goldstone boson octet $\pi$, $\eta$, $K$) and 
from four-baryon contact terms without and with two derivatives.
In the actual calculation, we utilize the recent $YN$ potential NLO19
established in Ref.~\cite{Haidenbauer:2019} and the original NLO interaction, 
denoted by NLO13, introduced in Ref.~\cite{Haidenbauer:2013}. 
The properties of these interactions are summarized selectively in the 
second and sixth column of Table~\ref{tab:T1}. 
The $\La p$ scattering lengths $a_s$ and $a_t$ in the
$^1S_0$ (singlet) and $^3S_1$ (triplet) partial waves are given together
with the $\chi^2$. The results in Table~\ref{tab:T1} correspond 
to a regulator with cutoff $\La = 600$~MeV, cf. Ref.~\cite{Haidenbauer:2013}
for details. A thorough comparison of the two versions NLO13 and NLO19 
for a range of cutoffs can be found in Ref.~\cite{Haidenbauer:2019}, 
where one can see that the two $YN$ interactions yield essentially 
equivalent results in the two-body sector. 
Note that the total $\chi^2$ is from a global fit to $36$ $\La N$ and $\Si N$ data 
points \cite{Haidenbauer:2019} while the $\chi^2$ for $\La p$ includes $12$ data 
points \cite{Sec68,Ale68}. In case of the data from Alexander et al.~\cite{Ale68}, 
set 2 from Table II of this paper is used where the momentum bins have been chosen so that there 
are roughly the same number of events per bin. The $\chi^2$ is calculated from the
central momentum. No averaging over the bin width is done in our calculations.
Both sets are shown in Fig.~\ref{fig:LN} together with the data by Sechi-Zorn et
al.~\cite{Sec68}. 

\begin{table*}[ht]
\caption{Properties of the considered $YN$ interactions.
 $\Lambda p$ singlet ($a_s$) and triplet ($a_t$) scattering lengths (in fm) 
 and the $\chi^2$ calculated based on different sets of data. The $\La$ 
 single particle potential $U_{\La}$ at $p_\La=0$ is given in MeV.  
  }
\label{tab:T1}
\vskip 0.1cm
\renewcommand{\arraystretch}{1.4}
\begin{center}
\begin{tabular}{|c|r|rrr|r|r|}
\hline
\hline
$YN$ interaction & NLO19 & Fit A & Fit B & Fit C & NLO13 & experiment \\
\hline
$a_s$                          & -2.91 & -4.00  & -4.50 & -5.00  & -2.91 & $-1.8 ^{+2.3}_{-4.2}$ \cite{Ale68}\\
$a_t$                          & -1.41 & -1.22  & -1.15 & -1.09  & -1.54  & $-1.6 ^{+1.1}_{-0.8}$ \cite{Ale68}\\
$\chi^2$ (total)               & 16.01 & 16.45  & 16.97 &  17.68 & 16.2 &\\
$\chi^2$ ($\La p$ only)        &  3.31 & 3.95   &  4.49 &  5.16  & 3.81 &\\
$\chi^2$ ($\Si^-p\to \La n$)   &  3.98 & 3.76   &  3.74 &  3.93  & 4.14 &\\
$U_\La(0)$                     &-32.6  &-31.7   &-31.3  & -30.8  & -21.6 &-27\,$\D$\,-30 \cite{Gal:2016}\\
\hline
\end{tabular}
\end{center}
\renewcommand{\arraystretch}{1.0}
\end{table*}

The binding energy of the hypertriton is much more sensitive to the strength of 
the $\La N$ interaction in the $^1S_0$ partial wave than to the one in the $^3S_1$
channel \cite{Miyagawa:1993,Gibson:1994}. This has been known for a long time and, 
e.g., has been implemented in form of the constraint $|a_s| \ge |a_t|$ in an attempt 
to determine the $\La p$ $S$-wave scattering length from their data by 
Alexander et al.~\cite{Ale68}. Faddeev calculations, say for the family of NSC97 potentials
\cite{Rijken:1999}, confirm that only $YN$ interactions where $|a_s|$ is significantly larger than 
$|a_t|$ lead to a bound hypertriton \cite{Nogga:2002}. 
Indeed, in the recent works by the J\"ulich-Bonn Group \cite{Haidenbauer:2005,Polinder:2006,Haidenbauer:2013,Haidenbauer:2019},
the empirical binding energy of the $^3_\Lambda {\rm H}$ was always considered as additional constraint when fixing the $YN$ interaction. 
Otherwise, it would have been impossible to pin down the relative
strength of the spin-singlet and spin-triplet 
$S$-wave contributions to the $\La p$ interaction, given the
complete absence of direct experimental information on the spin
dependence. 

It should be clear from the above discussion that we need to 
increase $|a_s|$ 
if we want to make the hypertriton more bound. And we have to reduce $|a_t|$ at the same time since we want to maintain the excellent overall description of $\La p$ and $\Si N$ scattering data. 
This can indeed be  achieved as documented in Table~\ref{tab:T1} where three illustrative fits
based on NLO19 are presented that produce the values $a_s = -4.0$~fm (A), $-4.5$~fm (B), and $-5.0$~fm (C),
respectively. As can be seen, the $\chi^2$ slowly deteriorates with increasing $|a_s|$.
However, overall, the variation is small and stays well within the one due to the inherent
regulator dependence of the employed EFT approach \cite{Haidenbauer:2013,Haidenbauer:2019}.
There is also practically no change in the in-medium properties as exemplified by the
value for the $\Lambda$ single-particle potential $U_\La (p_\La = 0)$, see 
Ref.~\cite{Haidenbauer:2015} for more information on the calculation. A comparison 
with the NLO13 interaction shows that off-shell properties of the interaction 
have a much larger impact on these in-medium properties \cite{Haidenbauer:2019}
than changes of the relative strength of the singlet and triplet interaction. 

The corresponding $\La p$ cross sections are shown in Fig.~\ref{fig:LN} by solid, 
dashed, and dash-dotted lines and one can see that the results 
are also visually well in line with the data. In the figure, we compare to the 
NLO19 interaction but the NLO13 results are almost indistinguishable
\cite{Haidenbauer:2019}. 

When using the hypertriton to constrain the relative 
strength of singlet and triplet interaction, we implicitly assume that 
$\La NN$ three-body forces (3BFs) only give a negligible contribution 
to the hypertriton binding energy. To support this assumption, we estimated 
effects from 3BFs in Ref.~\cite{Haidenbauer:2019} based on the underlying
power counting, the observed regulator dependence of the
$^3_\Lambda {\rm H}$ binding energy, and the actual magnitude of 
the effective 3BF mediated by an intermediate $\Si$. For the hypertriton, 
the estimate suggests that one should not expect more than $50$~keV 
from such forces in our framework where $\Sigma$s are explicitly taken 
into account.
 
\begin{table*}[ht]
\caption{$^3_\Lambda {\rm H}$ and $^4_\Lambda {\rm He}$ separation energies $E_\La$ (in MeV).
 The splitting $\Delta E_\La$ of the spin states of  $^4_\Lambda {\rm He}$  is also given. Cutoff values $\Lambda$ in the brackets are given in MeV.  
  }
\label{tab:T2}
\vskip 0.1cm
\renewcommand{\arraystretch}{1.3}
\begin{center}
\begin{tabular}{|l|c|cc|c|}
\hline
$YN$ interaction ($\Lambda$) & $E_\La$ ($^3_\Lambda {\rm H}$) 
& $E_\La$ ($^4_\Lambda {\rm He(0^+)}$) 
& $E_\La$ ($^4_\Lambda {\rm He(1^+)}$) 
& $\Delta E_\La$ ($^4_\Lambda {\rm He}$) \\
\hline
NLO19(500)    & 0.10  & 1.64 & 1.23 & 0.42 \\ 
NLO19(550)    & 0.09  & 1.54 & 1.24 & 0.30 \\ 
NLO19(600)    & 0.09  & 1.46 & 1.06 & 0.41 \\ 
NLO19(650)    & 0.10  & 1.53 & 0.92 & 0.61 \\ 
\hline 
Fit A (500)    & 0.32  & 2.11 & 1.14 & 0.97 \\ 
Fit A (550)    & 0.29  & 1.87 & 1.00 & 0.87\\ 
Fit A (600)    & 0.28  & 1.77 & 0.84 & 0.93 \\ 
Fit A (650)    & 0.29  & 1.83 & 0.67 & 1.16 \\ 
\hline
Fit B (500)    & 0.39  & 2.14 & 0.95 & 1.18 \\ 
Fit B (550)    & 0.38  & 2.00 & 0.93 & 1.06 \\ 
Fit B (600)    & 0.37  & 1.86 & 0.75 & 1.11 \\
Fit B (650)    & 0.36  & 1.89 & 0.57 & 1.32 \\ 
\hline 
Fit C (500)    & 0.47  & 2.24 & 0.89 & 1.36 \\ 
Fit C (550)    & 0.46  & 2.10 & 0.88 & 1.23 \\ 
Fit C (600)    & 0.44  & 1.92 & 0.68 & 1.24 \\
Fit C (650)    & 0.44  & 1.96 & 0.49 & 1.47 \\ 
\hline
NLO13(500)    & 0.14  & 1.71 & 0.79 & 0.92     \\ 
NLO13(550)    & 0.10  & 1.50 & 0.59 & 0.92 \\ 
NLO13(600)    & 0.09  & 1.48 & 0.58 & 0.90 \\  
NLO13(650)    & 0.09  & 1.49 & 0.62 & 0.88 \\ 
\hline
experiment  & $0.13(5)$ \cite{Davis:1992}  & $2.39(3)$ \cite{Juric:1973} & $0.98(3)$ \cite{Yamamoto:2015} & $1.406(2)(2)$ \cite{Yamamoto:2015}  \\
            & $0.41(12)$ \cite{Adam:2019}   &   &   &   \\
\hline
\end{tabular}
\end{center}
\renewcommand{\arraystretch}{1.0}
\end{table*}

In order to achieve a larger $|a_s|$ while preserving the 
good description of $YN$ 
data, we had to loosen the strict, self-imposed SU(3) symmetry for 
the contact interactions in the $\La N$ and $\Si N$ forces
\cite{Haidenbauer:2013,Haidenbauer:2019}. 
According to the SU(3) relations relevant for the scattering 
of two octet baryons \cite{Swart:1963,Dover:1991,Haidenbauer:2013},
the potentials in the
$^1S_0$ partial wave for $\La p \to \La p$ and $\Si^+ p \to \Si^+ p$ are both
dominated by the strength of the contact terms corresponding to the $\{27\}$ 
irreducible representation of SU(3). Since in the EFT interactions, 
but also in phenomenological $YN$ potentials \cite{Rijken:1999,Haidenbauer:2005},
the $^1S_0$ partial wave alone saturates basically the entire experimental
$\Si^+ p \to \Si^+ p$ cross section, cf. the discussion in \cite{Haidenbauer:2013},
there is no room for increasing the strength of the contact 
term in question in order to increase the $\La p$ scattering length.
It would immediately result in a drastic deterioration of the $\chi^2$. Therefore, in the present work, we kept the $\{27\}$ strength 
(i.e. the low-energy constant $\tilde C^{27}$ \cite{Haidenbauer:2019})
for $\Si^+ p \to \Si^+ p$ as determined in 
Ref.~\cite{Haidenbauer:2019} and varied only the corresponding 
contribution to the $\La p \to \La p$ channel. 
This introduces an SU(3) symmetry breaking in the leading-order 
contact terms, however, an SU(3) breaking that is 
well in line with chiral EFT and the associated 
power counting \cite{Haidenbauer:2013,Petschauer:2013}. 

Let us now come to the separation energies of light hypernuclei. 
As shown in previous calculations, the $\La$ separation energies 
are only mildly dependent on the underlying nucleon-nucleon ($NN$) interaction 
\cite{Nogga:2002,Haidenbauer:2019}. Therefore, we employ in all of 
the calculations shown here the same chiral semi-local momentum-space-regularized
$NN$ interaction of Ref.~\cite{Reinert:2017usi} at order N$^4$LO+ for a cutoff of $\La=450$~MeV.
We can expect that other $NN$ interactions will only lead to insignificant changes 
of the separation energies. The numerical accuracy of the $A=3$ and $A=4$ separation 
energies is better than $2$ and $20$~keV, respectively. For $A=5$ and $A=7$, the corresponding numerical accuracy is better than $40$ and $70$~keV. 

Whereas the contribution of 3BFs is probably negligible for $A=3$, 
it might become relevant for the more strongly bound $A=4$--$7$ systems.
This is supported by the results shown in Table~\ref{tab:T2}. 
The dependence of the separation energies on the regulator (cutoff) is an 
effect of next-to-next-to-leading order (N$^2$LO) which includes also 3BFs 
\cite{Hammer:2012id}. 
As expected, the variation is negligible for the hypertriton but 
can be as large as $200$--$300$~keV for $A=4$. For the discussion of 
$A=4$ separation energies, we have to take into account that this 
variation is a lower limit of our theoretical uncertainty. Even larger 
is the difference between the two different realizations of $YN$ interactions: NLO19 and 
NLO13. For the $1^+$ state, the predictions can differ as much as $500$~keV. 
Since both interactions predict very similar $YN$ phase shifts, this difference 
should be ultimately absorbed into similarly large 3BF contributions. 
Since the illustrative fits are based on the NLO19 parametrization, in the following, 
we will mostly compare with the NLO19 results \cite{Haidenbauer:2019}. 
This should show more clearly how a more attractive $\La N$ singlet interaction impacts 
binding energies of hypernuclei.

\begin{table*}[t]
\begin{center}
\caption{Separation energies of $^3_\Lambda {\rm H}$, 
$^4_\Lambda {\rm He}$, $^5_\Lambda {\rm He}$ and $^7_\Lambda {\rm Li}$
calculated with different $YN$ interactions that have been SRG evolved 
such that the $^5_\La$He separation energy is well reproduced. The corresponding 
SRG parameter $\La_{SRG}$ is given in fm$^{-1}$. 
}
\renewcommand{\arraystretch}{1.4}
\vskip 0.1cm
\label{tab:Li7sep}
{
\small  
\begin{tabular}{|l|r|r|rr|r|r|}
\hline
$YN$ interaction ($\Lambda$) & $\La_{SRG}$
& $E_\La$ ($^3_\Lambda {\rm H}$) 
& $E_\La$ ($^4_\Lambda {\rm He(0^+)}$) 
& $E_\La$ ($^4_\Lambda {\rm He(1^+)}$) 
& $E_\La$ ($^5_\Lambda {\rm He}$)
& $E_\La$ ($^7_\Lambda {\rm Li}$)\\
\hline
\hline
NLO19(500)     & 0.836 & 0.07 & 1.44 & 1.01 & $3.13(2)$ & $5.64(7)$ \\ 
NLO19(550)     & 0.806 & 0.07 & 1.33 & 0.94 & $3.12(2)$ & $5.61(6)$ \\ 
NLO19(600)     & 0.820 & 0.08 & 1.44 & 0.92 & $3.10(4)$ & $5.67(6)$ \\ 
NLO19(650)     & 0.868 & 0.11 & 1.71 & 0.91 & $3.14(2)$ & $5.86(5)$ \\ 
\hline 
Fit A (500)    & 0.849 & 0.23 & 1.75 & 0.95 & $3.11(2)$ & \\ 
Fit A (550)    & 0.832 & 0.24 & 1.70 & 0.83 & $3.12(2)$ & \\ 
Fit A (600)    & 0.836 & 0.27 & 1.84 & 0.80 & $3.14(2)$ & $6.09(4)$ \\ 
Fit A (650)    & 0.890 & 0.33 & 2.17 & 0.75 & $3.10(2)$ & \\ 
\hline
Fit B (500)    & 0.872 & 0.31 & 1.84 & 0.87 & $3.11(2)$ & \\ 
Fit B (550)    & 0.836 & 0.32 & 1.82 & 0.78 & $3.12(2)$ & \\ 
Fit B (600)    & 0.843 & 0.36 & 1.97 & 0.75 & $3.13(2)$ & $6.20(3)$ \\
Fit B (650)    & 0.910 & 0.42 & 2.31 & 0.70 & $3.14(2)$ & \\ 
\hline 
Fit C (500)    & 0.880 & 0.39 & 1.97 & 0.83 & $3.14(2)$ & \\ 
Fit C (550)    & 0.843 & 0.40 & 1.94 & 0.75 & $3.14(2)$ & \\ 
Fit C (600)    & 0.843 & 0.46 & 2.12 & 0.69 & $3.11(2)$ & $6.31(3)$ \\
Fit C (650)    & 0.913 & 0.51 & 2.41 & 0.65 & $3.11(2)$ & \\ 
\hline
NLO13(500)     & 0.868 & 0.11 & 1.69 & 0.98 & $3.16(2)$ & $5.86(5)$ \\ 
NLO13(550)     & 0.910 & 0.12 & 1.83 & 0.93 & $3.12(2)$ & $5.87(5)$ \\ 
NLO13(600)     & 0.910 & 0.13 & 1.94 & 0.94 & $3.11(2)$ & $5.89(5)$ \\  
NLO13(650)     & 0.912 & 0.13 & 1.98 & 0.93 & $3.14(2)$ & $5.96(5)$ \\ 
\hline
\hline
experiment  & -- & $0.13(5)$ \cite{Davis:1992} 
                 & $2.39(3)$ \cite{Juric:1973}     
                 & $0.98(3)$ \cite{Yamamoto:2015}     
                 & $3.12(2)$  \cite{Juric:1973} 
                 &  $5.58(3)$ \cite{Davis:1992} \\
            &    &  $0.41(12)$ \cite{Adam:2019}   
            &   
            &   
            &  
            &  $5.85(13)(10)$ \cite{Agnello:2009kt}  \\
\hline
\end{tabular}
}
\renewcommand{\arraystretch}{1.0}
\end{center}
\end{table*}

\begin{figure*}
\begin{center}
 \includegraphics[height=9cm]{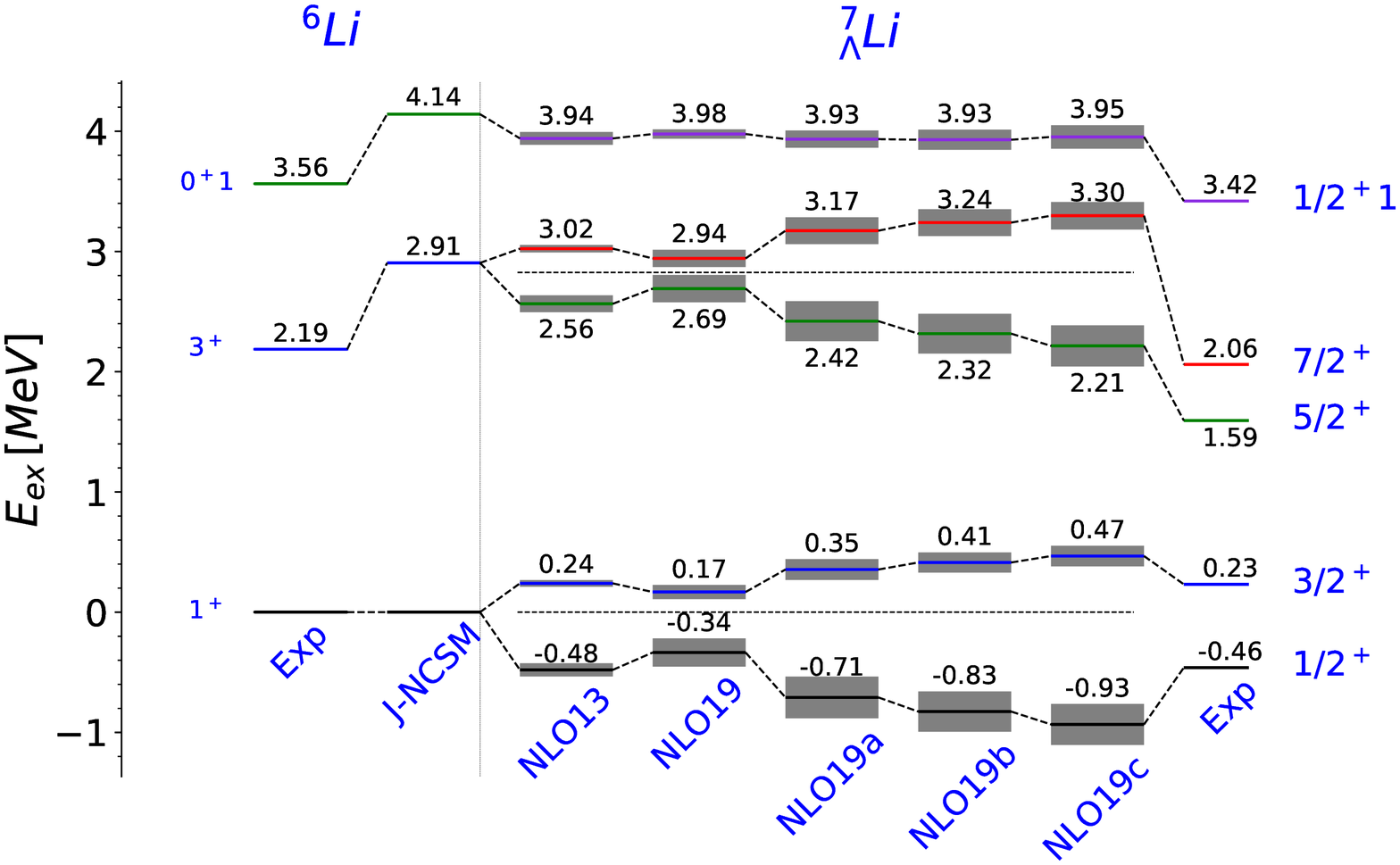}
\caption{Energy spectrum in $^7_\Lambda {\rm Li}$
calculated with different SRG evolved $YN$ interactions and compared to the 
spectrum of $^6$Li.The centroid energies of the first and second doublet are indicated by dashed lines. The grey bands show the dependence on the cutoff of the YN interaction. The interactions are defined in the text. \vspace{-0.7cm}}
\label{fig:Li7spect}
\end{center}
\end{figure*}

Comparing the energies for $^3_\Lambda {\rm H}$ for the different versions of the
$YN$ interaction NLO19, A, B, and C, one observes a  
dramatic increase of the $\Lambda$ separation energy. 
The prediction for fit B is already close to the 
STAR result based on their $^3_\Lambda {\rm H}$ events 
($0.35\pm 0.13$~MeV) and fit C even exceeds their combined
$^3_\Lambda {\rm H}$ + $^3_{\bar\Lambda} {\rm {\bar H}}$ value
of $0.41\pm 0.12$~MeV \cite{Adam:2019}.

There is also a noticeable change in the separation energies for the $0^+$ and $1^+$
states of $^4_\Lambda {\rm He}$. The value for the $0^+$ state becomes larger and 
is coming much closer to the empirical information with increasing singlet 
scattering length. Indeed, if one takes the latest result for the 
corresponding $^4_\Lambda {\rm H}$ binding energy as measure, 
$2.157\pm 0.005\pm 0.077$~MeV \cite{Schulz:2016kdc}, the 
results for Fit C already encompass the experimental value. 

The binding energy for the $1^+$ state decreases with increasing $|a_s|$. This is
not too surprising because, as argued in Refs.~\cite{Haidenbauer:2019,Gibson:1994},
this state is predominantly determined by the $\La p$ $^3S_1$ partial wave --
and the corresponding $|a_t|$ is reduced for Fit A to Fit C as compared to the 
reference $YN$ potential NLO19 \cite{Haidenbauer:2019}, see Table~\ref{tab:T1}.  
A remarkable feature of the results is that the splitting between the $0^+$ and 
$1^+$ states, $\Delta E_\La$ ($^4_\Lambda {\rm He}$), recently re-measured 
\cite{Yamamoto:2015} with very high accuracy, comes close to the empirical value. 
By contrast, the predictions of NLO19 fall short by more than a factor two for 
this quantity. The original NLO13 parameterization of the NLO interaction 
leads to somewhat larger splittings which amount to roughly $2/3$ of the 
experimental value.
These findings could indicate that 3BFs possibly play a significant role for 
this quantity. In any case, the large splitting measured in 
Ref.~\cite{Yamamoto:2015} certainly 
favors a somewhat increased singlet scattering length.

Finally, we present in Table~\ref{tab:Li7sep} and in Fig.~\ref{fig:Li7spect} our 
results for $^7_\La$Li. 
These results have been obtained by using similarity renormalization group (SRG) evolved 
$NN$ and $YN$ interactions and the Jacobi no-core shell model (J-NCSM).
For all of the calculations shown here, we again employ the semi-local momentum-space-regularized $NN$ interaction of Ref.~\cite{Reinert:2017usi} at order N$^4$LO+ for a cutoff of $\La=450$~MeV. The $NN$ interaction is evolved to a SRG flow parameter 
of 1.6~fm$^{-1}$.  It is well known that the 
separation energies of hypernuclei strongly dependent on the SRG flow parameter of the 
$YN$ interaction \cite{Wirth:2019}.
However, we found recently that the results for different SRG flow parameters 
are strongly correlated. In particular, it turned out that results are in 
good agreement with the ones for the original interactions 
once the flow parameter has been chosen such that one of the energies 
agrees with experiment \cite{Nogga:2019bwo}. 
We therefore choose the SRG parameter such that for each individual 
$YN$ interaction, the $^5_\La$He separation energy is reproduced. 
For this choice of SRG parameter, we find the $\Lambda$ separation energies 
given in Table~\ref{tab:Li7sep}. We also give the values of the chosen SRG parameters 
and results for the lighter 
systems where we can compare to the values obtained with bare interactions 
shown in Table~\ref{tab:T2}. For details of the calculations, 
we refer to Refs.~\cite{Liebig:2015kwa,Le:jncsm}.
We note in passing that, qualitatively, the weights of the contributions 
from the singlet and triplet $S$-wave $\La N$ interactions to the
$^5_\La$He binding energy are the same as for the $\La p$ cross section, 
see Eq.~(12) of Ref.~\cite{Haidenbauer:2019} or Sec.~5.2 of Ref.~\cite{Gibson:1994} for details.

By construction, we reproduce the separation energies for $^5_\La$He.
At the same time, we recover the predictions of the non-evolved 
interactions for $^3_\La$H at least within the theoretical uncertainty estimated 
by the cutoff variation. Also the changes of the $A=4$ separation energies 
due to the SRG-evolved interaction are within the bounds given by our 
3BF estimates from above. Interestingly, the predictions of NLO13 and NLO19 
are more similar to each other after the forces have been SRG-evolved. Especially, 
this holds for the predictions of the $1^+$ state. 

For $^7_\La$Li, the separation energy predictions for NLO13 and NLO19 are in
fair agreement with the experiments. However, the  values obtained 
with emulsion and counter experiments are somewhat different and the cutoff 
dependence indicates 3BF contributions of approximately $300$~keV. 

When employing the illustrative fits A, B and C, we recover the increased 
binding of $^3_\La$H and the $0^+$ state of $^4_\La$He and the decreased 
binding for the $1^+$ state of $^4_\La$He. Since the cutoff dependence 
for these fits follows the trend of the original NLO19 interaction for all light 
systems, we only calculated the separation energy for $^7_\La$Li 
for one cutoff for the modified interactions in order to save a substantial amount of computational resources. Although we find a visible increase of the 
separation energy with an increasing hypertriton energy, the 
overall changes are small compared to the expected 3BF contribution 
of $300$~keV. The modified interactions tend to overbind $^7_\La$Li. 
Nevertheless, the deviation from experiment is still comparable to 
possible 3BF contributions at least if one compares to the value of Ref.~\cite{Agnello:2009kt}. 

In Fig.~\ref{fig:Li7spect}, we summarize our results for the spectrum of $^7_\La$Li. Note that we do not reproduce the excitation spectrum of the $^6$Li core nucleus 
very well, because we neglect three-nucleon interactions in these calculations. 
Therefore, we focus our discussion on the relative positions of the levels of 
$^7_\La$Li and the corresponding $^6$Li core state for experiment and 
our predictions. 

Following Ref.~\cite{Gal:1978}, we introduce the centroid energy of a doublet by 
\begin{equation}
    \bar E = \frac{(J_N+1)}{2J_N+1}\ E_{+}
        +\frac{J_N}{2J_N+1}\ E_{-} \ .
\end{equation}
$E_{\pm}$ are the exitation energies of the $J_N\pm\frac{1}{2}$ 
state of the doublet where $J_N$ is the angular momentum of the corresponding core state. 

Shell-model studies  show that, for states related to only one core state, $\bar E$ will be independent of the spin-spin, tensor and hyperon spin-orbit $YN$ interaction \cite{Gal:1978,Millener:2010}. 
On the other side, the splitting of the 
two states will dependent on these contributions but will be insensitive to the nucleon spin orbit and the central $YN$ interaction. We expect that our J-NCSM calculations will reflect 
this behavior. These relations are not exact in our case since admixtures of the excited core states will always contribute. 

Because of this, it is instructive to plot the levels relative to 
$\bar E$ of the first $1/2^+$-$3/2^+$ doublet which is then at zero energy by construction as indicated by the dashed line. Interestingly, we observe for the second 
$5/2^+$-$7/2^+$ doublet that $\bar E =2.83$~MeV is independent of the interaction chosen. The energy is shown as the second dashed line.
The insensitivity of this energy to the chosen $YN$ 
interaction indicates
that the overall strength of the interactions is very similar. The 
different fits seem to be mostly different in their spin dependence. 
Therefore, we find that the doublet levels shift relative to the 
centroid energies and depend visibly on the interaction. 
The grey bands indicate the dependence of the results on the cutoff 
in the $YN$ interaction. 
One observes that there is a sizable cutoff dependence for most of the levels shown, indicating that 3BFs possibly affect the levels significantly. Also, NLO13 and NLO19 lead to slightly different predictions, further reinforcing 
that 3BFs are non-negligible for the excitation energies. 

Finally, we note that the $P$-wave interactions of all 
considered NLO forces are identical. We found that neglecting  
$P$- and higher partial waves in the interactions changes the
energies only marginally, well within
our cutoff dependence. 

All of the considered interactions qualitatively reproduce the experimental spectrum.
Quantitatively, however, none of the interactions is able to describe the experiment. 
For example, we find that the predicted $5/2^+$ state of $^7_\La$Li is located 
above the $3^+$ state of $^6$Li whereas the ordering is opposite for the 
experimental values.  
The splitting of the two lowest $^7_\La$Li states is correctly described by NLO13 
and NLO19. The illustrative fits A to C further increase the splitting bringing it 
away from the experimental value. But the deviations are mild if one 
considers possible 3BF contributions. 
In any case the result show that changes of the singlet scattering length also 
affect the spectra of $p$-shell hypernuclei. 
However, the changes are moderate and, therefore, the separation energy 
and spectrum remains qualitatively consistent with experiment for 
the illustrative fits. 

For completeness let us mention that in Ref.~\cite{Wirth:2019}
one can find a NCSM calculation for $^7_\La$Li based on the LO 
$YN$ interaction \cite{Polinder:2006} with cutoff $\Lambda=700$~MeV. 
Those results are qualitatively similar to our predictions
for NLO13 and NLO19 as far as the level ordering and splitting is 
concerned. However, they also reveal that there is a noticeable 
overall influence from the underlying $NN$ interaction, which 
makes a direct comparison difficult. 
We again stress that, in this work, we
restricted ourselves to NLO interactions to insure that all interactions are consistent with the available $YN$ scattering data. 

\section{Conclusions} 
Stimulated by the recent finding of the STAR Collaboration that 
the binding energy of the hypertriton could be significantly larger 
than so far assumed, 
we have investigated the consequences of a more strongly bound hypertriton for
the $\La p$ interaction and for the binding energies of the 
two $^4_\Lambda {\rm He}$ states. 
We have not found any principle reason that would speak against a larger 
$^3_\Lambda {\rm H}$ binding energy. The necessary increase of the attraction
in the $\La p$ $^1S_0$ partial wave is large but can be compensated by a 
correspondingly reduced attraction in the $^3S_1$ channel so that the overall
description of the $\La p$ and $\Si N$ data does not suffer.
The only caveat is that one has to give up strict SU(3) 
symmetry for the contact interactions in the $\La N$ and $\Si N$ channels.
However, such a symmetry breaking at the NLO level is anyway 
suggested by the 
counting scheme of SU(3) chiral EFT that we follow \cite{Petschauer:2013}. 
The improvements that we see in the predictions for the $^4_\Lambda {\rm He}$
binding energies certainly speak in favor of the scenario explored 
in this work. 

Using the Jacobi NCSM and SRG-evolved interactions, we 
extended the exploration to $^7_\La$Li. Increasing the 
hypertriton energy leads to an increased $^7_\La$Li 
separation energy. The changes are however small compared to possible 
effects from three-body forces. 

We also showed that the spectrum of $^7_\La$Li is affected by 
a change of the hypertriton binding energy or, respectively, the 
strength of the $\La p$ singlet interaction. 
In this case, the variations considered in the present study led to 
a slight deterioration in the description of the experimental spectrum. 
But given the significant uncertainties in the present predictions, 
these results do not really rule out a possibly more strongly bound 
hypertriton and/or a larger singlet scattering length. 
Shell-model calculations indicate that the spin-orbit 
interaction gives a sizable contribution to the excitation 
energies \cite{Gal:1978,Millener:2010}. This seems to be in contradiction to our observation that $YN$ $P$-waves 
do not contribute significantly. But the $P$-wave interactions are identical in all NLO forces, 
therefore, a more detailed  study including variations of the $\La p$ interaction in higher  partial waves is required to better understand this issue. 
Moreover, chiral \cite{Petschauer:2015elq} as well as SRG \cite{Wirth:2019} 3BFs, should be considered in order to reduce the theoretical uncertainties. 

In any case, our findings for the influence of an increased singlet scattering 
length on the hypernuclear binding and excitation energies should not be 
affected by such variations. 

Ultimately, there are two key quantities that can discriminate between the 
scenarios considered in the present study. One is the hypertriton binding 
energy itself.
Here improved measurements with noticeably reduced uncertainty
\cite{Achenbach:2018} would be extremely helpful. 
The other key quantity is the $\La p$ $^1S_0$ scattering length.
In principle, the latter could be extracted from studying the final-state 
interaction in reactions like $pp\to K^+\Lambda p$ 
\cite{Hinterberger:2004,Budzanowski:2010ib,Munzer:2018,Wilkin:2016}. 
However, the isolation of the spin-singlet amplitude requires a 
double-polarization experiment \cite{Gasparyan:2004}. 
Efforts at the COSY accelerator in J\"ulich to determine the strength of 
the spin-triplet $\La p$ interaction \cite{Hauenstein:2017}, where only 
single polarization is required, already suffered from low statistics 
and, unfortunately, did not provide robust results. 

Information on the $\La p$ scattering length can be also obtained from 
studying the $\La p$ correlation function measured in heavy-ion collisions 
or high-energetic $pp$ collisions \cite{Shapoval:2015,Cho:2017}.
There are already data from the STAR Collaboration \cite{Adams:2006} 
from a measurement in Au+Au collisions at $\sqrt{s}=200$~GeV 
and by the ALICE collaboration \cite{Acharya:2019}
in $pp$ collisions at $\sqrt{s}= 7$~TeV.
However, also here there is so far no detailed information on the
spin dependence and usually a purely statistical weight of the 
singlet and triplet states is assumed 
\cite{Shapoval:2015,Acharya:2019} . 
It would be rather important to find ways how to disentangle the 
spin states in those kind of experiments. 

\vskip 0.4cm
{\it Acknowledgments:}
We acknowledge stimulating discussions with Benjamin D\"onigus and
Josef Pochodzalla. 
This work is supported in part by the DFG and the NSFC through
funds provided to the Sino-German CRC 110 ``Symmetries and
the Emergence of Structure in QCD'' (DFG grant no. TRR~110; NSFC grant no. 11621131001)
and the VolkswagenStiftung (grant no. 93562).
The work of UGM was supported in part by The Chinese Academy
of Sciences (CAS) President's International Fellowship Initiative (PIFI)
(grant no.~2018DM0034). The numerical calculations were performed on JURECA
and the JURECA-Booster of the J\"ulich Supercomputing Centre, J\"ulich, Germany.


\end{document}